\documentclass[12pt]{article}
\usepackage{jheppub}
\usepackage{subcaption}
\usepackage{amsmath}
\usepackage{amssymb}

\usepackage{amsfonts}

\def\d{\operatorname{d}}

\newcommand{\nn}{\nonumber}

\def\bal#1\eal{\begin{align}#1\end{align}}

\def\secnum[#1]{\texorpdfstring{$#1$}{TEXT}}

\def\secnuml#1\secnumr{\texorpdfstring{$#1$}{TEXT}}

\def\eqa{\begin{eqnarray}}
\def\eqae{\end{eqnarray}}
\def\eq{\begin{equation}}
\def\eqe{\end{equation}}
\def\be{\begin{equation}}
\def\ee{\end{equation}}
\def\bea{\begin{eqnarray}}
\def\eea{\end{eqnarray}}
\def\ba{\begin{array}}
\def\ea{\end{array}}
\def\bd{\begin{displaymath}}
\def\ed{\end{displaymath}}

\def\>{\rangle}
\def\<{\langle}

\def\e{\epsilon}

\def\x{\xi}

\def\({\left(}
\def\){\right)}
\def\nn{\nonumber \\}

\title{Islands in Generalized Dilaton Theories}

\author{Jia Tian}
\affiliation{Kavli Institute for Theoretical Sciences (KITS),\\
	University of Chinese Academy of Sciences (UCAS), Beijing 100190, China }
\emailAdd{wukongjiaozi@ucas.ac.cn}
\date{}

\abstract{In this work we systematically study the island formula in the general asymptotically flat eternal black holes in generalized dilaton gravity theories or in higher dimensional spherical black holes. Under some reasonable and mild assumptions we prove that the island always appears barely outside of the horizon in the late time of Hawking radiation so that the information paradox is resolved. In particular, we find proper island in Liouville black hole which solves the puzzle of \cite{Li:2021lfo}.}\par

\begin{document}

\today

\maketitle

\section{Introduction}
Over the last five years there is a dramatic acceleration of progress on quantum aspects of black holes \cite{Bousso:2022ntt}. The most exciting achievement is the successful derivation of Page curve \cite{Page:1993wv} which resolves the long-standing information paradox \cite{Hawking:1975vcx} by proposing a new rule, the island formula \cite{Penington:2019npb,Almheiri:2019psf,Almheiri:2019hni}, for computing the entanglement entropy of the Hawking radiations. The island formula mimics  the quantum extremal surface (QES) prescription \cite{Engelhardt:2014gca} of the generalized entanglement entropy:
\bea \label{island}
S_R=\text{min}\Big\{\text{ext}\left[\frac{A(\partial I)}{4G_N}+S_{\text{semi-cl}}[\text{Rad}\cup I]\right]\Big\},
\eea 
where $I$ is the island which is a codimension-one region and $A(\partial I)$ is the area of its boundary $\partial I$, the QES. The surprising fact about this formula is that its right-hand side only depends on semi-classical physics and importantly the island is not added by hand but its existence can be justified by the replica tricks of the gravitational Euclidean path integral \cite{Penington:2019kki,Almheiri:2019qdq}. Another way to understand the island formula is by combining the AdS/BCFT correspondence and the brane world holography \cite{Sully:2020pza,Chen:2020uac,Chen:2020hmv,Suzuki:2022xwv}.\par 
The island formula has been successfully applied to black holes in various of gravitational theories \cite{Krishnan:2020fer,Caceres:2020jcn,Geng:2021wcq,Geng:2021iyq,Gautason:2020tmk,Hartman:2020swn,Hollowood:2020cou,Goto:2020wnk,Chen:2020jvn,Wang:2021mqq,Almheiri:2019psy,Hashimoto:2020cas,Wang:2021woy,Yu:2021cgi,Ahn:2021chg,Karananas:2020fwx,Lu:2021gmv,Krishnan:2020oun,Alishahiha:2020qza,Anegawa:2020ezn,He:2021mst}. However there is  also a notable counterexample \cite{Li:2021lfo} where it was claimed that the island formula can not save the information paradox of Liouville black hole. More puzzlingly this claim seems to be inconsistent with the systematic analysis of QES in general D-dimensional asymptotically flat (or AdS) eternal black hole performed in \cite{He:2021mst}.  In this work, we solve this puzzle by finding a new Liouville black hole solution with the help of the general construction of classical solutions of generalized dilaton theories. We also conduct a systematic analysis of islands in asymptotically flat eternal black holes. Our general results agree with the ones in \cite{He:2021mst}. 
\section{2D Generalized dilaton Gravity Theory}
Because of their remarkable solvability, 2D GDTs as toy models of quantum gravity serve as a laboratory for studying properties of black holes. The general action of 2D GDTs is 
\bea \label{2dg}
S_{}=\frac{1}{2\pi}\int \d^2 x\sqrt{-g}\left[X R-U(X)(\nabla X)^2-2 V(X)\right],
\eea 
which depends on the metric $g_{\mu\nu}$ and the scalar field $X$. The scalar field $X$ is usually related to the dilaton $\phi$ via $X=e^{-2\Phi}$. In terms of $\Phi$, the action is in a more familiar form
\bea \label{dil2}
S_{\text{dil}}=\frac{1}{2\pi}\int \d^2 x\sqrt{-g} e^{-2\Phi}\left[R-\tilde{U}(\Phi)(\nabla \Phi)^2-2\tilde{V}(\Phi)\right]
\eea 
with the identifications
\bea 
\tilde{U}=4 e^{-2\Phi} U,\quad \tilde{V}=e^{2\Phi}V.
\eea 
All the classical solutions of \eqref{2dg} can be found in a closed form with the help of the first-order formalism of GDT. We will show our convention and review the construction of classic solutions in the Appendix \ref{review}.  The general solution of \eqref{2dg} is given by\footnote{here we only consider the interesting linear dilaton vacua.} \cite{Grumiller:2002nm,Grumiller:2006rc}
\bea
\d s^2 &=& 2 e^Q \d v(\d X+(w(X)-C_0)\d v)\\
&=& 2\d v \d \tilde{X}+\xi(\tilde{X})\d v^2,
\eea 
where we have introduced
\bea 
&&\d \tilde{X}=\d X e^Q,\quad \xi(\tilde{X})=2e^Q(w-C_0), \label{xie}\\
&&Q=\int^X U(y)\d y, \label{Qfunc}\\
&&w=\int^X e^Q V(y) \d y \label{wfun}.
\eea 
The solution is parameterized by a constant $C_0$ which is usually related to the mass of the black hole. 
We can transform the metric to the diagonal gauge by introducing the coordinates $x^0=r, x^1=r$, thus $\d \tilde{X}=\dot{\tilde{X}} \d t+\tilde{X}' \d r$, $\d v=\dot{v} \d t+v' \d r$ and setting
\bea 
\tilde{X}=\tilde{X}(r),\quad \tilde{X}'+\xi v'=0.
\eea 
The resulting metric is 
\bea \label{conformalg}
\d s^2=\xi(\dot{v}^2 \d t^2-{v'}^2 \d r^2).
\eea 
If we further set $\tilde{X}=r$ and $\dot{v}=1$, we can get
\bea \label{schm}
\d s^2=\xi \d t^2-\frac{1}{\xi}\d r^2\, ,
\eea 
which is in the Schwarzschild gauge if the solution describes a black hole. In this paper we will focus on asymptotically flat solutions so we take the ansatz that  that  $\xi$ approaches to some constant
\bea 
\lim_{r\rightarrow \infty}\xi=-\xi_0^2,
\eea 
where $\xi_0^2$ should be positive to ensure the correct signature. Then we rescale the coordinates as
\bea 
t=\frac{t'}{\xi_0},\quad r=\xi_0 r'
\eea 
such that
\bea\label{schm}
&& \d s^2=\frac{\xi}{\xi_0^2}{\d t'}^2-\frac{\xi_0^2}{\xi}{\d r'}^2,\quad \lim_{r'\rightarrow \infty}\d s^2=-{\d t'}^2+{\d r'}^2.
\eea 
The horizon of the black hole is at $r_H',\quad \xi(r_H')=0$. The temperature and the entropy of the black hole are \cite{Iyer:1994ys}
\bea 
T_{\text{BH}}=\partial_{r'}{\xi/\xi_0^2}\,\Big|_{r'=r'_H},\quad S_{\text{BH}}=2X(r'_H).
\eea  
\section{General results}
\subsection{Setting up the calculation}
Our goal is to compute the entanglement entropy of the Hawking radiation of the eternal black hole with the island formula \eqref{island}.
For 2d GDT \eqref{2dg}, the first term is given by the value of dilaton field at the position of the island boundary $2X(\partial I)$ \cite{Iyer:1994ys}. $S_{\text{semi-cl}}[\text{Rad}\cup I]$ is semi-classical entanglement entropy of Hawking radiation in the region $\text{Rad}\cup I$. For eternal black holes, we first introduce Kruskal coordinates
\bea 
\d s^2=-e^{2\rho(y^+,y^-)}\d y^+ \d y^-
\eea 
 to cover the whole regions then we choose Rad to be two symmetric intervals $[y_{L\infty},y_{-a}]\cup [y_a,y_{R\infty}]$, where the coordinates of the end points are
\bea 
y_{L\infty}=(y_a^0,-\infty),\quad  y_{R\infty}=(y_a^0,\infty),\quad y_{-a}=(y_a^0,-y_a^1),\quad  y_{a}=(y_a^0,y_a^1).
\eea 
With this symmetric choice, it is reasonable to expect that the island\footnote{We  assume that the one-interval island  configuration dominates.} $[y_{-d},y_d]$ also enjoys this symmetry
\bea 
y_{-d}=(y_d^0,-y_d^1),\quad y_d=(y_d^0,y_d^1).
\eea 
 The Kruskal coordinates are related to two copies of Schwarzschild coordinates via
\bea \label{tranxy}
&&e^{l x_R^+}=l y_R^+,\quad e^{-l x_R^-}=-l y_R^-,\quad e^{-l x_L^+}=-l y_L^+,\quad e^{l x_L^-}=l y_L^-,\\
&&y^+_R\geq 0,\quad y^-_R\leq 0,\quad y^+_L\leq 0,\quad y^-_L\geq 0\,,
\eea
where $y_{R(L)}^\pm=y^0_{R(L)}\pm y^1_{R(L)}$ labels the position in the right (left) patch of Kruskal spacetime and $l$ is some convenient constant and
\bea\label{xtar}
x^\pm =t'\pm x^*,\quad x^*=\int \frac{\d r' }{-\xi/\xi_0^2}.
\eea 
From the transformations (about $x_R$) we find that
\bea 
&&l y^+ \d x^+ =\d y^+,\quad -l y^- \d x^- =\d y^-\, \\
&& e^{2\rho}=\frac{\xi(y^+ y^-)}{\xi_0^2l^2 y^+ y^-}.\label{e2r}
\eea 
We will model the Hawking radiation with a probe conformal field theory with a central charge $c<<1/G_N$ so can use the semi-classical formula to compute the entanglement entropy $S_{\text{semi-cl}}(\text{Rad}\cup I)$.

\subsection{Entanglement entropy without islands}
 In the early time of Hawking radiation, there are very few Hawking quanta in the interior of black holes so the island configuration can not be supported. So as more and more Hawking quanta escape to the infinity the asymptotic (Schwarzschild) observer should see a growing entanglement entropy. Assuming that the initial state of the quantum field is pure then
\bea 
S_{\text{semi-cl}}[\text{Rad}] &=&\frac{c}{6}\log\( {|(y_a-y_{-a})^+(y_a-y_{-a})^-| e^{\rho(y_a)}e^{\rho(y_{-a})}}\)\label{entropy1}\\
&=&\frac{c}{6}\log \(|2y_a^1|^2 e^{\rho(y_a^+,y_a^-)}e^{\rho(y_a^-,y_a^+)}\)\nonumber,\\
&=&\frac{c}{3}\log \(2\cosh(lt'_a)\)+\frac{c}{6}\log\(-\frac{\xi}{\xi_0^2l^2 }\),\label{ten}
\eea 
thus indeed as shown in the Page curve the  growth of early entanglement entropy is general. Without introducing islands, the entanglement entropy will exceed the Bekenstein-Hawking entropy before the black hole completely evaporates. However the Bekenstein-Hawking entropy should be the upper bound of entanglement entropy. This contradiction is the well known information paradox.  
\subsection{Entanglement entropy with islands}
In the late time of black hole evaporation, as we have seen that we have to include the islands. But as stressed in \cite{Almheiri:2020cfm}, the island region is not included by hand but its appearance is a result of evaluating gravitational path integral around non-trivial saddles. As a result, the black hole evaporation is unitary such that the entanglement entropy will vanish in the end as the Page curve shows. This implies the position of the $\partial I$ is very close to the horizon i.e. $y_d^-=y_d^0-y_d^1\approx 0$ which means that $y_d^1$ is very large. Therefore we can approximate the entanglement entropy $S_{\text{Rad}}([y_{-a},y_{-d}]\cup [y_{d},y_{a}] )$ with $2 S_{\text{Rad}}( [y_{d},y_{a}] )$ thus
\bea \label{entropy2}
S_{\text{island}}=\frac{4}{G_N} X(y_d) +\frac{c}{3}\log \( |(y_a-y_d)^+(y_a-y_d)^-|e^{\rho(y_a)}e^{\rho(y_d)}\),
\eea 
where we have added back the Newton's constant $G_N$ to indicate that in the semi-classical limit, the first term should be much larger than the second term. Taking this approximation as an ansatz and then solving $y_d$ by extremizing the generalized entanglement entropy \eqref{entropy2} we can show indeed this approximation is correct. \par 
Differentiating \eqref{entropy2} with respect to $y_d^\pm$ gives two extremal conditions
\bea 
&&\frac{4}{G_N}\frac{\d X}{\d y_d^+}+\frac{c}{3}\(\frac{1}{y_d^+-y_a^+}+\frac{\d \rho}{\d y_d^+}\)=0,\label{de1}\\
&&\frac{4}{G_N}\frac{\d X}{\d y_d^-}+\frac{c}{3}\(\frac{1}{y_d^--y_a^-}+\frac{\d \rho}{\d y_d^-}\)=0.\label{de2}
\eea 
Recalling that $\tilde{X}=r$  first term can be evaluated as
\bea 
\frac{\d X}{\d d_R^\pm}=\frac{\d X}{\d \tilde{X}}\frac{\d r}{\d y_d^\pm}=e^{-Q}\frac{\xi_0 \d r'}{\d y_d^\pm}=-e^{-Q}\frac{\xi \d x^*}{\xi_0 \d y_d^\pm}=-e^{-Q}\frac{\xi}{2l \xi_0 y_d^\pm}
\eea 
where $\xi(z)$ should be understood as function of $y_d^+y_d^-\equiv z$. Using the expression \eqref{e2r} the last term can be computed as
\bea 
\frac{\d \rho}{\d y_d^\pm}=\frac{1}{2 y_d^\pm}\(\frac{\xi'}{\xi}\frac{\d z}{y_d^\pm}-\frac{1}{y_d^\pm}\)=\frac{1}{2 y_d^\pm}\(\frac{\xi'z}{\xi}-1\)
\eea 
Then the two equations \eqref{de1} and \eqref{de2} can be written as
\bea \label{sse}
&&\frac{1}{3}\frac{1}{y_d^\pm-y_a^\pm}-\frac{2\xi e^{-Q}}{ cG_N l \xi_0}\frac{1}{y_d^\pm}+\frac{1}{3}\frac{\xi-z \xi'}{2  \xi}\frac{1}{y_d^\pm}=0,
\eea 
from which we can obtain the relation
\bea 
&&\frac{y_d^+}{y_a^+-y_d^+}=\frac{y_d^-}{y_a^--y_d^-},\quad \rightarrow\quad y_a^+ y_d^-=y_a^- y_d^+,\quad \text{or}\quad \frac{y_a^+}{y_a^-}=\frac{y_d^+}{y_d^-}. \label{ratio}
\eea 
In the semi-classical limit, $c G_N<<1$ thus the last term in \eqref{sse} can be neglected so we only need consider the equations
\bea 
&&\frac{z}{y_d^- y_a^+-z}=-\frac{6\xi e^{-Q}}{cG_Nl\xi_0},\quad \frac{z}{y_d^+ y_a^--z}=-\frac{6\xi e^{-Q}}{cG_Nl\xi_0}
\eea 
from which we can derive an \textit{single} equation for $z$
\bea \label{eqz}
z\(1-\frac{\epsilon}{\xi e^{-Q}}\)^2=y_a^+y_a^-,\quad \epsilon\equiv \frac{cG_N l \xi_0}{6},
\eea 
where we will also assume that the ''effective" coupling $\epsilon$ is small.
Solving it and using the relation\eqref{ratio} we can solve $y_d^\pm$.\par 
In the end we need to transfer to the Schwarzschild  coordinate because it describes the asymptotic observer and it turns out it is more convenient to do the computation with the Schwarzschild coordinates. Let the positions  $y_a$ and $y_d$ are $(t_a,a)$ and $(t_d,d)$ in the Schwarzschild coordinates such that
\bea 
&&\frac{1}{l}Ye^{l t_a}=y_a^+,\quad \frac{1}{l}Ye^{-l t_a}=-y_a^-,\quad Y=e^{l y^*},\quad y^*=\int^a \frac{\d r'}{-\xi/\xi_0^2},\\
&&\frac{1}{l}De^{l t_d}=y_d^+,\quad \frac{1}{l}De^{-l t_d}=-y_d^-,\quad D=e^{l d^*},\quad d^*=\int^d \frac{\d r'}{-\xi/\xi_0^2},
\eea 
thus the entanglement entropy \eqref{entropy2} can be written as
\bea 
S_{\text{island}}=\frac{4X(d)}{G_N}+\frac{c}{3}\log\frac{1}{l^2}\(Y^2+D^2-2YD\cosh l(t_a-t_d)\)+\frac{c}{3}\(\rho(d)+\rho(a)\).
\eea 
Varying with respect to $t_d$ implies $t_d=t_a$ therefore the extremal of $S_{island}$ is time-independent as expected.
Assuming the temperature of the black hole \eqref{schm} is not zero so it can evaporate then $\xi$ has a single zero at $r'=r'_H$. It implies that $d^*$ has a logarithm singularity at $d=r_H'$ so we can rewrite 
\bea \label{dstar}
d^*=f(d)+\frac{1}{2l}\log(d-r'_H)
\eea 
where $f(d)$ is regular at $r'_H$. Since we have assumed that the island is very close to the horizon $d\approx r'_H$ or equivalently $z\approx 0$ thus we have the approximation
\bea 
&&e^{2l d^*}=e^{2l f(d)}(d-r'_H)=-\frac{1}{l^2}z,\quad \rightarrow \\
&& d=r'_H-e^{-2l f(r'_H)}\frac{1}{l^2}z+\mathcal{O}(z^2).
\eea  
 If we also assume that $e^{-Q}$ is regular and not vanishing at $d=r'_H$ then 
\bea 
\xi e^{-Q}=(d-r'_H)g(d),
\eea 
where $g(d)$ is a regular at $r'_H$. Thus \eqref{eqz} becomes
\bea 
z(1-\frac{\epsilon l^2 e^{2l f(r'_H)}}{z g(d)})^2\approx z+\frac{1}{z}\(\frac{\epsilon l^2 e^{2l f(r'_H)}}{ g(r'_H)}\)^2-2\frac{\epsilon l^2 e^{2l f(r'_H)}}{g(r'_H)}=y_a^+y_a^-\equiv y^2.
\eea 
There is indeed one solution which satisfies our ansatz $z\approx 0$: 
\bea 
&&z=\frac{\beta^2}{y^2}+\mathcal{O}(\beta^3),\quad \beta=\frac{\epsilon l^2 e^{2l f(r'_H)}}{g(r'_H)},\\
&&y_d^+=-\frac{\beta}{y_a^-},\quad y_d^-=-\frac{\beta}{y_a^+}\, ,
\eea
which leads to 
\bea 
S_{\text{island}}&=&\frac{4 X(r'_H)}{G_N}+\frac{c}{3}\log\(|y_a^+y_a^-| e^{\rho(y_a^\pm)}\),\\
&=&2S_{\text{BH}}+S_{\text{matter}}
\eea 
where $S_{\text{matter}}$ is the quantum correction of order $\mathcal{O}(G_N^0)$ due to the presence of matter fields. This is the main result of this paper: under some reasonable assumptions for a general asymptotically flat eternal black hole in GDT we can find island such that (generalized) entanglement entropy of Hawking radiation follows Page curve which resolves the information paradox.  
 
\section*{Examples}
\section{CGHS model}
The most well-studied GDT which admits an asymptotically flat black hole solution is the Callan-Giddings-Harvery-Strominger model (CGHS model) \cite{Callan:1992rs}. Islands in this model have been found in \cite{Gautason:2020tmk,Anegawa:2020ezn}. In this section, we will rederive the island with our general procedures to confirm the validity of our general analysis. The action of the CGHS model is 
\bea\label{CGHSa}
S=\frac{1}{2\pi}\int \d^2x\sqrt{-g}\left[e^{-2\phi}(R+4(\nabla \phi)^2+4\lambda^2)\right]
\eea .
\subsection{The geometry}
Comparing \eqref{CGHSa} with \eqref{dil2}
we recognize that
\bea 
U=-\frac{1}{X},\quad V=-2\lambda^2 X\, .
\eea 
Therefore according to our general discussion, we can compute the following data 
\bea
&&e^Q=\frac{1}{X},\quad X=\exp{\tilde{X}},\quad w=-2\lambda^2 X=-2\lambda^2 e^{\tilde{X}}\\ 
&&\xi=-2C_0 e^{-\tilde{X}}-4\lambda^2,\quad \xi_0=2\lambda.
\eea 
So the Schwarzschild metric is 
\bea \label{s1}
\d s^2=-\(\frac{C_0}{2\lambda^2}e^{-2\lambda r' }+1\){\d t'}^2+\frac{{\d r'}^2}{\(\frac{C_0}{2\lambda^2}e^{-2\lambda r' }+1\)}.
\eea 
It is easy to find that the horizon and curvature singularity are located at
\bea 
r'_H=-\frac{1}{2\lambda}\log\(-\frac{2\lambda^2}{C_0}\),\quad r'_s=-\infty.
\eea 
Therefore we should take $\lambda^2>0$ and $C_0\equiv-C<0$ to get the black hole geometry. To summarize, the classical solution describes a asymptotically flat black hole with metric and dilaton
\bea \label{sch2}
&&\d^2 s=-(1-\frac{C}{2\lambda^2}e^{-2\lambda r'})\d^2 t'+\frac{1}{(1-\frac{C}{2\lambda^2}e^{-2\lambda r'})}\d^2 r',\\
&&X=\exp\(2\lambda r'\). \label{dil}
\eea 
The temperature of this black hole can be found using the equation
\bea 
T=-\frac{1}{4\pi}\partial_{r'} \sqrt{-\frac{g_{t't'}}{g_{r'r'}}}\,|_{r=r_H}=\frac{\lambda}{2\pi},
\eea 
and the entropy is given by the Wald formula
\bea 
S=2X |_{r'=r'_H}=2e^{2\lambda r'}|_{r'=r'_H}=\frac{C}{\lambda^2}.
\eea 
Introducing the new variable \eqref{xtar}
\bea 
x^*=\int \frac{\d r'}{1-\frac{C}{2\lambda^2}e^{-2\lambda r'}}=\frac{\log\(2\lambda^2 e^{2 r'\lambda}-C\)}{2\lambda },\eea 
we can obtain the Kruskal coordinates $y^\pm$ as for example through
\bea \label{tr1}
e^{\lambda x^+}=\lambda y^+,\quad e^{-\lambda x^-}=-\lambda y^-
\eea 
thus the metric and dilaton become
\bea \label{kal}
&&\d^2 s=-\frac{\d y^+\d y^-}{C-\lambda^2 y^+ y^-},\quad e^{2\rho}=\frac{1}{C-\lambda^2 y^+ y^-},\\
&&X=\frac{1}{2\lambda^2}\(C-\lambda^2 y^+y^-\).
\eea 
In the Kruskal coordinates, the horizon is located at $y^+y^-=0$ and the singularity is located at $y^+y^-=C$.
As we have shown in the general discussion, without including the island the entanglement entropy is given by the general formula \eqref{ten}. Let us focus on the derivation of the island.
\subsection{The derivation of island}
In the late time, the entanglement entropy with island is given by:
\bea 
S_{\text{island}}=\frac{1}{G_N} \frac{4}{2\lambda^2}\(C- \lambda^2y_d^+ y_d^- \) +\frac{c}{3}\log \(\frac{ |(y_a-y_d)^+(y_a-y_d)^-|}{\sqrt{C-\lambda^2 y_a^+ y_a^-}\sqrt{C-\lambda^2 y_d^+ y_d^-}}\).
\eea 
Taking derivative with respect to $y_d^-$ and $y_d^+$ we obtain the equations
\bea 
&&\frac{\lambda^2 y_d^+}{6(C-\lambda^2 y_d^+y_d^-)}+\frac{1}{3(y_d^--y_a^-)}-\frac{2 y_d^+}{c G_N }=0,\\
&&\frac{\lambda^2 y_d^-}{6(C-\lambda^2 y_d^+y_d^-)}+\frac{1}{3(y_d^+-y_a^+)}-\frac{2 y_d^-}{c G_N }=0,
\eea 
The exact solutions can be straightforwardly obtained but the exact solutions are very complicated. To exact the useful information we again take the semi-classical limit $G_N\rightarrow 0$. In this limit, we find that the non-trivial solutions are
\bea \label{ds1p}
y_d^+=-\frac{c G_N }{6y_a^-} ,\quad y_d^-=-\frac{c G_N }{6y_a^+}
\eea 
and corresponding extremal entanglement entropy in the Schwarzschild coordinate is
\bea 
S_{\text{island}}=2\frac{1}{G_N}\frac{C}{\lambda^2}+\frac{c}{3}\log \(\frac{| y_a^+y_a^-|}{\sqrt{C\(C-\lambda^2 y_a^+y_a^-\)}}\)
\eea 
which is time-independent and coincides with the results in \cite{Anegawa:2020ezn,Gautason:2020tmk}.\par 
Alternatively, we can apply our general result \eqref{eqz}:
\bea 
z(1-\frac{\epsilon}{\xi e^{-Q}})^2=y^2\quad \rightarrow \quad z(1-\frac{\epsilon}{2\lambda^2 z})^2=y^2,\quad \epsilon=\frac{cG_N \lambda^2}{3}
\eea 
which has two solutions 
\bea 
z_1=y^2+\frac{\epsilon}{\lambda^2}-\frac{\epsilon^2 }{4\lambda^4 y^2}+\mathcal{O}(\epsilon^3),\quad z_2=\frac{\epsilon^2}{4 y^2 \lambda^4}+\mathcal{O}(\epsilon^3).
\eea 
The solution $z_1$ leads to the trivial solution while the solution $z_2$ leads to \eqref{ds1p}.
\section{Liouville gravity}
A particular generalization of CGHS model is the one with exponential potential. The action is 
\bea 
S=\frac{1}{2\pi} \int d^2 x\sqrt{-g}\left[RX+\sum_i 4\alpha_i^2e^{\beta_i X} \right].
\eea 
There are some interesting reasons to consider such exponential potentials. It is shown in \cite{Cruz:1997nj} that this kind of model admits extra (conformal) symmetries. If we add $2X$ in the potential, this kind of models as deformations of JT gravity is shown to have a matrix model dual \cite{Witten:2020wvy}. So it means that
\bea \label{liouvillep}
U(X)=0,\quad V(X)=-2\sum_i \alpha_i^2e^{\beta_i X}\, .
\eea 
For simplicity, let us take $k=1$ and the model is called the Liouville gravity.  Surprisingly, it is claimed in \cite{Li:2021lfo} that island formula can \textit{not} resolve the information paradox of Liouville gravity based on the black hole solution which is found in \cite{Cruz:1997nj,Mann:1993rf}. In this section, we will use the general solution of GDT to derive a different solution such that the island formula successfully resolves the information paradox.
\subsection{The geometry}
Given the potentials \eqref{liouvillep} we can compute the following data
\bea 
&& Q=0,\quad X=\tilde{X}=r,\quad w=-\frac{2\alpha^2 e^{\beta X}}{\beta},\\
&& \xi=2(w-C_0^L)=-2(\frac{2\alpha^2 e^{\beta X}}{\beta}+C_0^L),\quad  \xi_0=\sqrt{2C_0^L},
\eea 
where we choose $\beta<0$. The corresponding Schwarzschild metric is
\bea \label{sch3}
\d s^2=-(1+\frac{2\alpha^2 e^{\beta \sqrt{2C_0^L} r'}}{C_0^L\beta}){\d t'}^2+\frac{1}{1+\frac{2\alpha^2 e^{\beta \sqrt{2C_0^L}r'}}{ C_0^L\beta}}{\d r'}^2.
\eea 
Therefore the horizon is at 
\bea 
r'_H=\frac{1}{\beta}\log\(\frac{-C_0^L \beta}{2\alpha^2}\),
\eea 
and the Ricci scalar is 
\bea 
R=-4\beta e^{\beta \sqrt{2C_0^L}r'}\alpha^2.
\eea 
So to ensure asymptotic flatness we can set
\bea 
\beta<0,\quad C_0^L>0,\quad \alpha^2>0,
\eea 
Note that the metric \eqref{sch3} is same as \eqref{sch2} if we identity
\bea 
\beta=-\frac{2\lambda}{\sqrt{2C_0^L}},\quad \alpha^2=\frac{C }{2\lambda}\sqrt{\frac{C_0^L}{2}}.
\eea 
If we also set $C_0^L=2\lambda^2 $ such that $\alpha^2=\frac{C}{2},\beta=-1$ then the dilaton is given by
\bea 
X=\tilde{X}=2\lambda r',
\eea 
thus the dilaton of the Liouville black hole is the Logarithm of the one of the CGHS black hole. Below we will keep $C_0^L$ general. Similarly the geometry in the Kruskal coordinates are\footnote{We have double checked that the solutions indeed solve the equations of motion in the second order formalism. }
\bea \label{liukru}
&&\d^2 s=-\frac{\d y^+\d y^-}{C-\lambda^2 y^+ y^-},\\
&&X=-\frac{1}{\beta}\log\left[\frac{1}{2\lambda^2}\(C-\lambda^2 y^+y^-\)\right].
\eea 
Since the geometry is same the entanglement entropy without island is also same as one in CGHS black hole. Let us focus on the entanglement entropy of the Hawking radiation in the presence of possible islands. 
\subsection{The derivation of island}
In the late time, the entanglement entropy with island is given by:
\bea 
S_{\text{island}}=\frac{4}{G_N}\frac{\sqrt{2C_0^L}}{2\lambda }\log\left[ \frac{1}{2\lambda^2}\(C-\lambda^2 y_d^+ y_d^- \)\right] +\frac{c}{3}\log \(\frac{ |(y_a-y_d)^+(y_a-y_d)^-|}{\sqrt{C-\lambda^2 y_a^+ y_a^-}\sqrt{C-\lambda^2 y_d^+ y_d^-}}\). \nn
\eea 
The extremal conditions are
\bea 
&&-\frac{4}{G_N}\frac{\sqrt{2C_0^L}}{2\lambda }\frac{\lambda^2y_d^+}{C-\lambda^2y_d^+y_d^-}+\frac{c}{3(y_d^--y_a^-)}+\frac{c y_d^+}{6(C-\lambda^2y_d^+y_d^-)}=0,\\
&&-\frac{4}{G_N}\frac{\sqrt{2C_0^L}}{2\lambda }\frac{\lambda^2y_d^-}{C-\lambda^2y_d^+y_d^-}+\frac{c}{3(y_d^+-y_a^+)}+\frac{c y_d^-}{6(C-\lambda^2y_d^+y_d^-)}=0\,.
\eea 
These equations are quadratic so can be easily solved. In the limit $G_N \rightarrow 0$ the two solutions behave as
\bea 
&& y_d^-=\frac{cG_N C\beta}{12\lambda^2 y_a^+}=-\frac{c G_N C}{6\sqrt{2 C_0^L} y_a^+\lambda},\quad y_d^+=\frac{cG_N C\beta}{12\lambda^2 y_a^-}=-\frac{c G_N C}{6\sqrt{2 C_0^L} y^-_a\lambda},\label{phy}\\
&&y_d^-=y^-_a+\mathcal{O}(G_N),\quad y_d^+=y^+_a+\mathcal{O}(G_N),\label{nonphy}.
\eea 
Let us try to derive these solutions directly from our general result \eqref{eqz}:
\bea \label{eqzz}
z(1-\frac{\epsilon}{\xi e^{-Q}})^2=y^2-\quad \rightarrow \quad z\((1+\frac{\epsilon}{2C_0^L})-\frac{\epsilon C}{2C_0^L\lambda^2 z}\)^2=y^2, 
\eea 
with
\bea 
\epsilon=\frac{cG_N \lambda \sqrt{2 C_0^L}}{6}.
\eea 
The equation \eqref{eqzz} is also quadratic with solutions to be
\bea 
&&z_1=\frac{C^2\epsilon^2}{4{C_0^L}^2\lambda^4 y^2}+\mathcal{O}(\epsilon^3),\quad z_2=y^2-\frac{\epsilon(\lambda^2y^2-C)}{C_0^L \lambda^2}+\mathcal{O}(\epsilon^2)
\eea 
which will correspond to \eqref{phy} and \eqref{nonphy}, respectively.
The first solution \eqref{phy} is the non-trivial one which gives the generalized entanglement entropy \bea 
S_{\text{island}}=2\frac{1}{G_N}\frac{\sqrt{2 C_0^L}}{\lambda}\log\(\frac{C}{2\lambda^2}\)+\frac{c}{3}\log \(\frac{ |y_a^+y_a^-|}{\sqrt{C\(C-\lambda^2 y_a^+y_a^-\)}}\).
\eea 
Thus we have derived the Page curve for the Liouville black hole. The reason why we succeed is that we have derived another black hole solution whose parameters are opposite to those in the solutions used in \cite{Li:2021lfo} or derived in \cite{Cruz:1997nj,Mann:1993rf}. Let us revisit the black geometry which is used in \cite{Li:2021lfo}. 
\subsection{The other black geometry}
To get that solution we start from \eqref{sch3} and reverse the radial coordinate
\bea 
r'\rightarrow -r',
\eea 
such that the metric becomes
\bea 
\d s^2=-(1+\frac{2\alpha^2 e^{-\beta \sqrt{2C_0^L} r'}}{C_0^L\beta}){\d t'}^2+\frac{1}{1+\frac{2\alpha^2 e^{-\beta \sqrt{2C_0^L}r'}}{ C_0^L\beta}}{\d r'}^2.
\eea 
The position of the event horizon and the Ricci scalar are
\bea 
r_H'=\frac{1}{\sqrt{2C_0^L}\beta}\log\(-\frac{2\alpha^2}{C_0^L\beta}\),\quad R=-4\beta \alpha^2 e^{-\sqrt{2C_0^L}\beta r'}.
\eea 
So requiring the asymptotic flatness at $r'\rightarrow \infty$ forces the choice 
\bea 
\beta>0, 
\eea 
and having a well-defined horizon forces the choice
\bea 
\alpha^2<0.
\eea 
With these choices, the solution in the Kruskal coordinates are still given by \eqref{liukru}
\bea 
&&\d^2 s=-\frac{\d y^+\d y^-}{C-\lambda^2 y^+ y^-},\\
&&X=-\frac{1}{\beta}\log\left[\frac{1}{2\lambda^2}\(C-\lambda^2 y^+y^-\)\right].
\eea 
but with different identification 
\bea 
\beta=\frac{2\lambda}{\sqrt{2C_0^L}},\quad \alpha^2=-\frac{C}{2\lambda}\sqrt{\frac{C_0^L}{2}}.
\eea 
But in this black hole solution the position of the island is at
\bea
&& y_d^-=\frac{cG_N C\beta}{12\lambda^2 y_a^+}=\frac{c G_N C}{6\sqrt{2 C_0^L} y_a^+\lambda},\quad d^+=\frac{cG_N C\beta}{12\lambda^2 y_a^-}=\frac{c G_N C}{6\sqrt{2 C_0^L} y_a^-\lambda},\label{phywrong}
\eea
which is in the left Kruskal patch. This contradicts the assumption that $y_d$ is in the right patch and this is why \cite{Li:2021lfo} claims the failure of island formula. However we have shown this is only because a ''wrong" solution is used. \par 
To summarize, island can \textbf{\textit{save}} the information paradox of Liouville gravity.
\section{ab-family}
In this section, we consider a large family of dilaton gravity theories which has the following  potentials 
\bea 
U(X)=-\frac{a}{X},\quad V(X)=-\frac{B}{2}X^{a+b}.
\eea 
In general, there are two free parameters and sometimes this family is called the ab-family \cite{Katanaev:1996ni}.
From our general discussion, the classical solution is 
\bea 
\d^2 s=2 X^{-a}\d X \d v-X^{-2}(2 C_0+\frac{B X^{b+1}}{b+1})\d^2 v.
\eea 
Because we are interested in the asymptotically flat black hole solutions $\lim_{X\rightarrow \infty} R=0$ we will choose \cite{Katanaev:1996ni} 
\bea 
b=a-1,\quad a\in(0,1).
\eea  
This choice can be understood form the behavior of Ricci scalar
\footnote{here we consider the case $b\neq 1$.}
\bea 
R=-2a C_0 X^{a-2}+\frac{bB(a-b-1)}{b+1}X^{a+b-1},
\eea 
by noticing that $C_0$ is related to mass of black hole therefore the solution with $C_0=0$ should be the Minkowski spacetime. Following the general discussion we compute 
\bea 
&&Q=-a\log X,\quad X=(1-a)^{\frac{1}{1-a}}\tilde{X}^{\frac{a}{1-a}},\quad w=-\frac{B}{2a}(1-a)^{\frac{a}{1-a}}\tilde{X}^{\frac{a}{1-a}},\\
&& \xi=-\frac{B}{a}-\frac{2C_0^f}{(1-a)^{\frac{a}{1-a}}\tilde{X}^{\frac{a}{1-a}}},\quad \xi_0=\sqrt{\frac{B}{a}}.
\eea 
Thus the corresponding Schwarzschild metric is 
\bea 
\d s^2=-(1-\frac{1}{2\lambda{r'}^{\frac{a}{1-a}}}){\d t'}^2+\frac{1}{1-\frac{1}{2\lambda{r'}^{\frac{a}{1-a}}}}{\d r'}^2,
\eea 
where 
\bea 
\frac{1}{2\lambda}=-\frac{2C_0^f a}{B}\(\sqrt{\frac{B}{a}}(1-a)\)^{\frac{a}{a-1}},\quad C_0^f<0,\quad \lambda>0.
\eea 
So the horizon is located at
\bea 
r'_H=(2\lambda)^{\frac{a-1}{a}}. 
\eea 
Next we can transform it the conformal gauge by introducing 
\bea 
x^*=\int \frac{\d r'}{1-\frac{1}{2\lambda{r'}^{\frac{a}{1-a}}}}=2 (a-1) \lambda  {r'}^{\frac{1}{1-a}} \, _2F_1\left(1,\frac{1}{a};1+\frac{1}{a};2 {r'}^{\frac{a}{1-a}} \lambda \right)+c_1,
\eea 
where $c_1$ is a constant which can be chosen for our convenience. The hypergeometric function generally can not be inverted to write $r'(x^*)$ as a function of $x^*$ explicitly. However for the special case of  $a=1/2$, we can invert the function with product logarithm:
\bea 
&& x^*=r'+\frac{\log(2\lambda r'-1)}{2\lambda}+c_1,\quad \quad \rightarrow \\
&& r'=\frac{1}{2\lambda}+\frac{W_0\left(e^{2\lambda (x^*-c_1)-1}\right)}{2\lambda },
\eea 
where $W_0$ is the principle branch of the Lambert $W$ function or product logarithm. 
Thus it is natural to introduce the Kruskal coordinates as
\bea 
&&e^{\pm \lambda x^\pm}=\pm \lambda y^\pm,\quad c_1=-\frac{1}{2\lambda},\\
&&\d s^2=-e^{2\rho}\d y^+ \d y^-,\quad e^{2\rho}=\frac{1}{e^{W_0(-\lambda^2 y^2)}-\lambda^2 y^2},\\
&&X=\sqrt{\frac{B}{2}}\frac{1}{4\lambda}\(1+W_0(-\lambda^2 y^2)\)
,\quad y^2=y^+y^-.
\eea 
The equation for determining the position of island becomes
\bea 
z \left(\frac{\epsilon 2^{1/4}  \sqrt{\lambda } \sqrt{W_0\left(\lambda ^2 (-z)\right)+1}}{B^{5/4} W_0\left(\lambda ^2 (-z)\right)}+1\right)^2=y^2,\quad \epsilon=\frac{c G_N \lambda}{6}\sqrt{\frac{B}{a}}.
\eea 
Assuming $\epsilon \rightarrow 0$ we can expand the left-hand side to the first order of $z$ then the we will obtain two solutions 
\bea 
z_1=y^2+\frac{2^{1/4}\alpha (2-3 y^2 \lambda^2)}{B^{5/4}\lambda^{3/2}}+\mathcal{O}(\epsilon^2),\quad z_2=\frac{\beta^2}{y^2},\quad \beta^2=\frac{\sqrt{2}\epsilon^2}{B^{5/2}\lambda^3 }.
\eea 
Thus the physical solution is
\bea 
y_d^+=-\frac{\beta}{y_a^-},\quad y_d^-=-\frac{\beta}{y_a^+}\, ,
\eea 
which leads to 
\bea 
S_{\text{island}}=\sqrt{\frac{B}{2}}\frac{1}{G_N \lambda}+\frac{c}{3}\log \frac{|y_a^+y_a^-|}{\sqrt{e^{W_0(-\lambda^2 y^2)}-\lambda^2 y^2}}+\mathcal{O}(G_N).
\eea 
For generic $a$, we observe that $x^*$ can be always decomposed into 
\bea 
x^*=f(r')+\frac{1-a}{a}(2\lambda)^{\frac{a-1}{a}}\log(r'-r'_H),
\eea 
where $f(r')$ is regular at $r'=r'_H$ as we expect in \eqref{dstar}. Here we omit the further analysis since it is very similar to the general result.

\section{Reissner-Nordstrom}
Islands in charged black hole have been studied in \cite{Wang:2021woy,Karananas:2020fwx,Ahn:2021chg,Yu:2021cgi}. Even though they considered 4-dimensional black holes, effectively and technically the model is still 2-dimensional after a dimensional reduction of the two sphere. Therefore we can also study them with our general procedure. To support a Reissner-Nordstrom black hole, the simplest choice of  potentials are
\bea 
U(X)=-\frac{1}{2X},\quad V(X)=-\lambda^2+\frac{A}{X}
\eea 
which lead to the following data
\bea 
&&e^Q=\frac{1}{\sqrt{X}},\quad X=\frac{\tilde{X}^2}{4},\quad w=-\frac{2(A+\lambda^2 X)}{\sqrt{X}}=-\lambda^2\tilde{X}-\frac{4 A}{\tilde{X}},\label{ee1}\\
&&\xi=-4\lambda^2-\frac{4C_0^{R}}{\tilde{X}}-\frac{16 A}{\tilde{X}^2},\quad \x_0=2\lambda.\label{ee2}
\eea 
Thus the metric and dilaton are
\bea 
&&\d s^2=-(1+\frac{C_0^R}{2 r' \lambda^3}+\frac{A}{r'\lambda^4}){\d t'}^2+\frac{1}{1+\frac{C_0^R}{2 r' \lambda^3}+\frac{A}{r'\lambda^4}}{\d r'}^2,\\
&& X=\lambda^2{r'}^2.
\eea 
Comparing with the standard Reissner-Nordstrom we can identify the following parameters
\bea 
&&C_0^R=-4M \lambda^3,\quad A=\lambda^4 Q_c^2,\\
&&M=\frac{r_++r_-}{2},\quad Q_c=\sqrt{r_+ r_-},
\eea 
where $M$ and $Q_c$ are the mass and charge of the black hole and $r_\pm$ are positions of the outer $(+)$ and inner $(-)$ horizons. Next we introduce new variable
\bea 
&&x^*=\int \frac{\d r'}{1-\frac{2M}{r'}+\frac{Q_c^2}{{r'}^2}}=r'+\frac{r_+^2\log(r'-r_+)-r_-^2\log(r'-r_-^2)}{r_+-r_-},\\
&&\exp(2x^*)=e^{2r'}(r'-r_+)^{\frac{2r_+^2}{r_+-r_-}}(r'-r_-)^{-\frac{2r_-^2}{r_+-r_-}},\label{e2star}
\eea 
to get the conformal metric
\bea 
\d s^2=-H(r')\d x^+ \d x^-,\quad H(r')=1-\frac{2M}{r'}+\frac{Q_c^2}{{r'}^2}\, .
\eea 
The Kruskal coordinates can be defined as
\bea 
&& e^{lx^+}=ly^+,\quad e^{-lx^-}=-ly^-,\quad l=\frac{r_+-r_-}{2r_+^2}.
\eea 
In order to use \eqref{eqz} to solve the position of the island, we first express $d$ in terms of $z$. Using \eqref{e2star} we can directly get
\bea 
-l^2 z=e^{2d l}(d-r_+)(d-r_-)^{-\frac{r_-^2}{r_+^2}},
\eea 
which leads to
\bea \label{rp}
d=r_+-e^{-2l r_+}l^2(r_+-r_-)^{\frac{r_-^2}{r_+^2}}z+\mathcal{O}(z^2)\,.
\eea 
Recall that we expect that $z\rightarrow 0$. Substituting \eqref{rp} into \eqref{eqz} with help of \eqref{ee1} and \eqref{ee2} we end up an equation of $z$. The non-trivial solution is
\bea 
&&z=\frac{\beta^2}{y^2},\quad \beta^2=\frac{\epsilon^2e^{4l r_+}r_+^2(r_+-r_-)^{-\frac{2r_-^2}{r_+^2}-2}}{16 l^4 \lambda^6 y^2},\quad \epsilon=\frac{cG_Nl\lambda}{3},\\
&&y_d^+=-\frac{\beta}{y_a^-},\quad y_d^-=-\frac{\beta}{y_a^+}\, .
\eea

\subsection{Other Charged dilaton Black Hole I}
We can also consider island in other charged dilaton black hole. In \cite{Ahn:2021chg}, the charged dilaton black hole has the metric
\bea 
\d s^2=-r^2\(1-\frac{2M}{r^2}+\frac{Q_c^2}{4 r^4}\)\d t^2+\(1-\frac{2M}{r^2}+\frac{Q_c^2}{4 r^4}\)^{-1}\d r^2+r^2(\d x^2+d y^2).
\eea  
The effective 2D model is
\bea 
\d s^2=-H(r)\d t^2+r^2 H(r)^{-1}\d r^2,\quad X=r^2,
\eea 
where 
\bea 
H(r)=r^2\(1-\frac{2M}{r^2}+\frac{Q_c^2}{4 r^4}\).
\eea 
To transform to the Schwarzschild metric let us introduce
\bea 
dr'= 2 r\d r,\quad \rightarrow,\quad r'=r^2,
\eea 
such that 
\bea \label{d1}
&&\d s^2=-H(r'){\d t'}^2+H(r')^{-1}{\d r'}^2,\quad t'=t/2,\\
&&H(r')=4r'-8M-\frac{Q_c^2}{r'}=\frac{4(r-r_+)(r-r_-)}{r'},\quad r_\pm=M\pm\frac{\sqrt{4M^2-Q_c^2}}{2}.
\eea 
The geometry is asymptotically flat. The outer event horizon and curvature singularity are located at $r'=r_+$ and $r'=0$ respectively. The solution can be embedded into dilaton gravity by choosing the possible potentials to be
\bea 
U(X)=0,\quad V(X)=-2+\frac{Q_c^2}{2 X^2},\quad C_0=4M.
\eea 
When $Q_c=0$, the geometry \eqref{d1} reduces to Rindler patch. From the potential we obtain
\bea 
Q=0,\quad X=\tilde{X}=r',\quad  w=-2 X-\frac{Q_c^2}{2 X},\quad \xi=8M-\frac{Q_c^2}{X}-4X.
\eea 
Next we introduce new variable
\bea 
&&x^*=\int \frac{\d r'}{H(r')}=\frac{r_+}{4(r_+-r_-)}\log(r'-r_+)-\frac{r_-}{4(r_+-r_-)}\log(r'-r_-),\\
&&e^{2l x^*}=(r'-r_-)^{-\frac{r_-}{r_+}}(r'-r_+),\quad \frac{1}{2l}=\frac{r_+}{4(r_+-r_-)}.
\eea  
Following the general procedure we find that in the late time the position of the island is
\bea 
&&d=r_+-l^2(r_+-r_-)^{\frac{r_-}{r_+}}z,\\
&&z=\frac{\beta^2}{y^2},\quad \beta^2=\frac{\epsilon^2 r_+^2(r_+-r_-)^{-\frac{2(r_-+r_+)}{r_+}}}{16 l^4},\quad \epsilon=\frac{cG_N l}{6},\\
&&y_d^+=-\frac{\beta}{y_a^-},\quad y_d^-=-\frac{\beta}{y_a^+}\, ,
\eea 
which coincides with the results in \cite{Ahn:2021chg}.
\subsection{Other Charged dilaton Black Hole II}
In \cite{Yu:2021cgi}, the charged dilaton black hole has the metric
\bea 
\d s^2=-W(r)\d t^2+W^{-1}\d r^2+R(r)^2\d\Omega^2,
\eea 
with the function
\bea 
W(r)=\(1-\frac{r_+}{r}\)\(1-\frac{r_-}{r}\)^n,\quad R^2=r^2\(1-\frac{r_-}{r}\)^{1-n},\quad n\in[0,1).
\eea 
The effective 2D model is
\bea 
\d s^2=-W(r)\d t^2+W^{-1}\d r^2,\quad X=r^2\(1-\frac{r_-}{r}\)^{1-n}\equiv f(r).
\eea 
The corresponding 2d dilaton potentials can be
\bea 
&&e^Q=\frac{\d f^{-1}(X)}{\d X}\equiv {f^{-1}}',\quad U(X)=\frac{\d \ln({f^{-1}}')}{\d X},\label{v1}\\
&&V(X)=-\frac{1}{2}e^{-Q(X)}\frac{\d \(e^{-Q(X)}W(f^{-1}(X))\)}{\d X}\label{v2}.
\eea 
In general, $f(r)$ is hard to invert but to solve the island the  explicit expressions of the potentials are not needed. We only need the following quantity
\bea 
\xi e^{-Q}=-W \frac{d X}{\d r}=-\frac{(2r-(1+n)r_-)(r-r_+)}{r},
\eea 
which appears in \eqref{eqz} and the relation between $r$ and $z$:
\bea \label{xstar2}
x^*&=&\int \frac{\d r}{W(r)}=\frac{r^n}{(r-r_-)^{n-1}}+(nr_-+r_+)B_{1-\frac{r_-}{r}}(1-n,0) \nonumber \\
&-&\(\frac{r_+}{r_+-r_-}\)^nB_{t}(1-n,0),\quad t=\frac{r_+}{r_+-r_-}\(1-\frac{r_-}{r}\),
\eea 
where $B_\alpha(a,b)$ is the incomplete beta function. Note that
\bea 
\lim_{t\rightarrow 1}B_{t}(1-n,0)=-\log(t-1)
\eea 
thus let us denote $x^*$ as
\bea 
&&x^*=\mathcal{R}+\(\frac{r_+}{r_+-r_-}\)^n\log(r-r_+), \\
&&e^{2l x^*}=e^{2l\mathcal{R}}(r-r_+),\quad \frac{1}{2l}=\(\frac{r_+}{r_+-r_-}\)^n.
\eea 
It implies that
\bea 
d=r_+-{l^2}e^{-2l\mathcal{R}(r_+)}z+\mathcal{O}(z^2).
\eea 
Substituting into \eqref{eqz} we can solve
\bea 
&& z=\frac{\beta^2}{y^2},\quad \beta=\frac{\epsilon r_+}{e^{-2l\mathcal{R}(r_+) }(2r_+-r_--nr_-)},\quad \epsilon=\frac{cG_N }{6}, \\
&&y_d^+=-\frac{\beta}{y_a^-},\quad y_d^-=-\frac{\beta}{y_a^+}\, ,
\eea 
which is consistent with the result in \cite{Yu:2021cgi} while our method is much simpler.
\section{Kaluza-Klein black holes}
Our last example is the 4-dimensional Kaluza-Klein black hole. The island of this black hole is studied in \cite{Lu:2021gmv}. The metric of a non-rotating KK black hole in 4d asymptotically flat spacetime is 
\bea \label{KKBH}
\d s^2=-W(r)\d t^2+\frac{\d r^2}{W(r)}+H^{1/2}r^2\d \Omega^2,
\eea 
where
\bea 
W(r)=\frac{f(r)}{\sqrt{H(r)}},\quad f(r)=1-\frac{r_h}{r},\quad H(r)=1+\frac{Q_c}{r}.
\eea 
When $Q=0$, \eqref{KKBH} is just the 4 dimensional Schwarzschild black hole and the corresponding dilaton gravity model has the potentials
\bea 
U(X)=-\frac{1}{2X},\quad V(X)=-\lambda^2.
\eea  
To embed the solution \eqref{KKBH} into dilaton gravity we can first identify 
\bea 
X=H^{1/2}r^2\equiv f(\tilde{X}),\quad r=\tilde{X}.
\eea 
Supposing that $f(\tilde{X})$ is invertible, we can solve $\tilde{X}=f^{-1}(X)$. The potentials of the corresponding GDT can be found with \eqref{v1} and \eqref{v2}.
For the solution \eqref{KKBH}, the results are very involved.  The results in  leading order of $Q$ are
\bea 
 U(X)=-\frac{1}{2X}+\frac{3 Q^2}{32 X^2},\quad V(X)=-1+\frac{r_h Q}{4 X}+\frac{Q^2(2\sqrt{X}-3r_h)}{8 X^{3/2}}.
\eea 
To compute the generalized entanglement entropy we can directly use the 2d dilaton gravity solutions 
\bea 
&&\d s^2=-W(r)\d t^2+\frac{\d r^2}{W(r)},\\
&&X=\sqrt{1+\frac{Q_c}{r}}r^2,\quad W(r)=(1-\frac{r_h}{r})(1+\frac{Q_c}{r})^{-\frac{1}{2}},
\eea 
which are already in the Schwarzschild coordinates. To derive the island first we compute
\bea 
\xi e^{-Q}=-W \frac{d X}{\d r}=-\frac{(4r+3Q_c)}{2(r+Q_c)}(r-r_h),
\eea 
and
\bea 
x^*&=&\int \frac{\d r}{W(r)}=\sqrt{r(Q_c+r)}+(Q_c+r_h)\sinh ^{-1}\left(\ \sqrt{\frac{r}{Q_c}} \right)\\
&-&\sqrt{r_h(Q_c+r)}\log\(1+\sqrt{\frac{{r(Q_c+r_h)}}{{r_h(Q_c+r)}}}\)+\sqrt{r_h(Q_c+r)}\log\(1-\sqrt{\frac{{r(Q_c+r_h)}}{{r_h(Q_c+r)}}}\)\nonumber \\
&=&\mathcal{R}+\sqrt{r_h(Q_c+r)}\log(r-r_h),
\eea 
where $\mathcal{R}$ is again regular at $r=r_h$. Then we can solve $d$ in terms of $z$:
\bea 
&&e^{2l x^*}=e^{2l \mathcal{R}}(r-r_h),\quad \frac{1}{2l}=\sqrt{r_h(Q_c+r)},\\
&&d=r_h-l^2e^{-2l \mathcal{R}(r_h)}z.
\eea 
Substituting into \eqref{eqz} we find
\bea 
&& z=\frac{\beta^2}{y^2},\quad \beta=\frac{2 \epsilon (r_h+Q_c)}{e^{-2l\mathcal{R}(r_h) }(3 Q_c + 4 rh)},\quad \epsilon=\frac{cG_N }{6}, \\
&&y_d^+=-\frac{\beta}{y_a^-},\quad y_d^-=-\frac{\beta}{y_a^+}\, .
\eea 
\section{Conclusion and Discussion}
In this work we have studied island formula \eqref{island} in the general asymptotically flat eternal black holes in GDT. Under some reasonable and mild assumptions we prove that the island always appears barely outside of the horizon in the late time of Hawking radiation so that the information paradox is resolve, in particular, in the Liouville gravity theory in which it was reported in \cite{Li:2021lfo} that the island proposal failed. We find that failure is due to the use of a ``wrong" black hole solution. With the help of general construction of classical solutions of GDT we find a different black hole solution where the island appears as expected. We further apply our general analysis to a large family of GDT and several 4-dimensional black holes including different charged dilaton black holes and the KK black hole. It turns out that our procedure for finding island is much simpler.\par 
There are some possible generalizations of our analysis.
\begin{itemize}
	\item Our general analysis should be simply generalized to the asymptotically AdS black holes in GDT by gluing a flat bath. Since after gluing the flat bath, the whole spacetime is similar to the asymptotically flat black hole and cut-off surface $y_a$ can be chosen to be boundary of the AdS space. 
	\item In this work we only consider the classical solutions of GDT. It is also possible to include the quantum effect which comes from the conformal anomaly following for example \cite{Hartman:2020swn}. 
	\item  It is also possible to generalize our results to single-sided black hole and consider a truly evaporating black hole. Some examples are \cite{Hartman:2020swn,Wang:2021mqq}.
\end{itemize}

\begin{acknowledgments}
I would like to thank Yi-Jun He, Yang An and many of the members of KITS for interesting related discussions. JT is supported by the National Youth Fund No.12105289 and the internal fund of KITS.
\end{acknowledgments}

\appendix
\section{Review of 2D GDT}
\label{review}
\subsection*{Conventions} 
The local Lorentz metric and the Lorentz transformation invariant tensor are chosen to be 
\bea 
\eta_{ab}=\eta^{ab}=\begin{pmatrix}
	-1&0\\
	0&1
\end{pmatrix},\quad \epsilon^a_{~b}=\begin{pmatrix}
0&1\\1&0	
\end{pmatrix}.
\eea 
Thus the Levi-Civita tensors are
\bea 
\epsilon^{ac}=\epsilon^a_{~b}\eta^{bc}=\begin{pmatrix}
	0&1\\
	-1&0
\end{pmatrix},\quad \epsilon_{ab}=\eta_{ac}\epsilon^c_{~b}=\begin{pmatrix}
	0&-1\\
	1&0
\end{pmatrix}.
\eea 
The volume form is related to the local Lorentz basis $e^a$ via
\bea 
\epsilon &=&\frac{1}{2}\epsilon_{ab}e^a\wedge e^b=\frac{1}{2}\epsilon_{ab}e^a_{~\mu}e^b_{~\nu}dx^\mu \wedge dx^\nu=\frac{1}{2}\epsilon_{ab}\(e^a_{~1}e^b_{~0}-e^a_{~0}e^b_{~1} \)\d x^1\wedge \d x^0\\
 &=& \(e^1_{~1}e^0_{~0}-e^1_{~0}e^0_{~1} \)\d x^1\wedge \d x^0=\sqrt{-g}\d x^1\wedge \d x^0\rightarrow  \sqrt{-g}\d^2 x.
\eea 
In 2d the spin connection should be proportional to $\epsilon$: $\omega^{a}_{~b}=\omega \epsilon^{a}_{~b}$ and Ricci tensor two-form is then given by
\bea 
R_{ab}&=&\d \omega \epsilon_{ab},\quad R_{ab}=\frac{1}{2}(R_{\mu\nu})_{ab}dx^\mu \wedge dx^\nu,\quad (R_{\mu\nu})_{ab}=\epsilon_{ab}\(\partial_\mu\omega_\nu-\partial_\nu\omega_\mu\).
\eea 
So the Ricci scalar is 
\bea 
(R_{\mu\nu})_{ab} e^{a\mu }e^{b \nu}&=&\epsilon^{\mu\nu}(\partial_\mu\omega_\nu-\partial_\nu\omega_\mu)=2\epsilon^{\mu\nu}\partial_\mu \omega_{\nu}=2|e|^{-1}\tilde{\epsilon}^{\mu\nu}\partial_\mu \omega_{\nu} \\
&=&2|e|^{-1}(\partial_0\omega_1-\partial_1 \omega_0)
\eea 
On the other hand we have
\bea 
\d\omega=\partial_\mu \omega_{\nu}dx^\mu \wedge dx^\nu=(\partial_1 \omega_0-\partial_0 \omega_1)dx^1 \wedge dx^0=-\frac{1}{2} R\sqrt{-g}\d^2 x.
\eea 
The torsion two-form is given by
\bea 
T^a=(D)^a_{~b}e^b&=&(\delta^a_{~b}\d +\omega^a_{~b})e^b=\d e^a+\omega^a_{~b}\wedge e^b,
\eea 
with its components are
\bea \label{torsion}
T^a_{\mu\nu}&=&\partial_\mu e^a_{\nu}-\partial_\nu e^a_{\mu}+(\omega_\mu)^a_{~b}e^b_\nu-(\omega_\nu)^a_{~b}e^b_\mu \\
&=&D_\mu e^a_\nu-D_\nu e^a_\mu.
\eea 
It is convenient to use the Light-cone gauge:
\bea 
x^\pm=\frac{1}{\sqrt{2}}(x^0\pm x^1),\quad x^0=\frac{1}{\sqrt{2}}(x^++x^-),\quad x^1=\frac{1}{\sqrt{2}}(x^+-x^-).
\eea 
The Lorentz transformation connecting these gauges is
\bea 
\Lambda^a_{~\bar{a}}=\frac{1}{\sqrt{2}}\begin{pmatrix}
	1&1\\
	1&-1
\end{pmatrix}.
\eea 
Thus we can find that 
\bea 
\eta_{\bar{a}\bar{b}}=\begin{pmatrix}
	0&-1\\
	-1&0
\end{pmatrix},\quad \epsilon_{\bar{a}\bar{b}}=\begin{pmatrix}
	0&1\\
	-1&0
\end{pmatrix},\quad \epsilon^{\bar{a}}_{~\bar{b}}=\begin{pmatrix}
	1&0\\
	0&-1
\end{pmatrix}
\eea 
such that the torsion form \eqref{torsion} can be expressed as
\bea 
T^\pm=(\d\pm \omega)e^{\pm}.
\eea 
\subsection{The first order formalism of GDT}
The action \eqref{2dg} is equivalent to  
\bea 
I_{\text{gen}}[e_a,\omega,X,X^a]= \int \(X \d\omega+X_a(\d e^a+\epsilon^a_{~b}\omega \wedge e^b)+\frac{1}{2}\e^{ab}e_a\wedge e_b \mathcal{V}(X,X^cX_c)\). \nonumber \\ 
\eea 
We will first solve all its classical solution and then prove the equivalence.
Varying with respect to $\omega$ gives
\bea 
&&X \d \delta \omega +X^a \epsilon_{ab}\delta \omega\wedge e^b=-\d X\wedge \delta \omega-X^a\epsilon_{ab}e^b\wedge \delta \omega \rightarrow \nonumber \\
&& \d X+X^a\epsilon_{ab}e^b=0. \label{e1}
\eea 
Varying with respect to $e$ we get
\bea 
&& X^a\d \delta e_a+X^a \epsilon_{a}^{~b}\omega \wedge \delta e_b+\frac{1}{2}\epsilon^{ab}\(\delta e_a\wedge e_b-e_a\wedge \delta e_b\)\mathcal{V} \rightarrow \nonumber \\
&&-\d X^a\wedge \delta e_a+X^b \epsilon_{b}^{~a}\omega \wedge \delta e_a-\epsilon^{ab}e_b\wedge \delta e_a \mathcal{V} \rightarrow \nonumber \\
&& dX^a-X^b \epsilon_{b}^{~a}\omega+\epsilon^{ab}e_b\mathcal{V}=dX^a+X^b \epsilon_{~b}^{a}\omega+\epsilon^{ab}e_b\mathcal{V}=0,\label{e2}
\eea 
where in the last line we used $\epsilon_{b}^{~a}=\eta_{bc}\epsilon^c_{~d}\eta^{da}=-\epsilon_{~b}^{a}$.
The other two equations of motion are
\bea 
&&\d \omega+\frac{1}{2}\epsilon^{ab}e_a\wedge e_b \frac{\partial \mathcal{V}}{\partial X}=0,\label{e3}\\
&&\d e_a+\epsilon_a^{~b}\omega\wedge e_b+\frac{1}{2}\epsilon^{ab}e_a\wedge e_b \frac{\partial \mathcal{V}}{\partial X^a} ,\label{e4}.
\eea 
In the light-cone gauge the equations of motion become
\bea 
&&\d X+X^+e^--X^- e^+=0,\label{e37} \\
&&(\d\pm \omega)X^\pm\pm\mathcal{V}e^\pm=0,\label{e38}\\
&&\d\omega+\epsilon \frac{\partial V}{\partial X}=0,\label{e39}\\
&&(\d\pm \omega)e^\pm+\epsilon \frac{\partial \mathcal{V}}{\partial X_\pm}=(\d\pm \omega)e^\pm-\epsilon \frac{\partial \mathcal{V}}{\partial X^\mp}=0,\label{e310}
\eea 
where the volume form is $\epsilon=e^+\wedge e^-$ and in the last line we have used $X_\pm=-X^\mp$.
From \eqref{e38} we get
\bea 
X^-\d X^++X^+\d X^-+\mathcal{V}(X^- e^+-X^+ e^-)=0,
\eea 
then using \eqref{e37} we get
\bea \label{conserve}
\d(X^-X^+)+\mathcal{V}(X^-X^+,X)\d X=0.
\eea 
This equation indicates that these exists a conserved quantity defined by integrating \eqref{conserve}. \par 
If $X^+\neq 0$, from \eqref{e38} we can get
\bea 
\omega=-\frac{\d X^+}{X^+}-Z\mathcal{V},\quad Z\equiv\frac{e^+}{X^+},
\eea 
and from \eqref{e37} we can get
\bea 
e^-=-\frac{\d X}{X^+}+X^-Z.
\eea 
Substituting the expression of  volume form
\bea 
\e=\frac{1}{2}\epsilon_{ab}e^a\wedge e^b=e^+\wedge e^-=\d X\wedge Z
\eea 
into \eqref{e310} gives
\bea 
&&\d e^++\omega \wedge e^+-\d X\wedge Z \frac{\partial \mathcal{V}}{\partial X^-}=0\\
&&=X^+ \d Z+\d X^+\wedge Z-\d X^+\wedge Z- \d X \wedge Z\frac{\partial \mathcal{V}}{\partial X^-}=0.
\eea 
Therefore we end up with
\bea \label{eqdz}
\d Z=-\frac{Z\wedge \d X}{X^+}\frac{\partial \mathcal{V}}{\partial X^-}.
\eea 
Taking the ansatz of $Z$ as
\bea 
Z=\d v e^{Q(X)},\quad \d Z=e^{Q(X)}\frac{\d Q}{\d X}\d X\wedge dv\, ,
\eea 
and substituting into \eqref{eqdz} gives
\bea 
&&\frac{\d Q}{\d X}=\frac{1}{X^+}\frac{\partial \mathcal{V}}{\partial X^-},\rightarrow \\
&&Q=\int^X \frac{1}{X^+}\frac{\partial \mathcal{V}}{\partial X^-}. \label{q}
\eea 
Recall that the metric is 
\bea 
\d s^2=\eta_{ab}e^a e^b=-2e^+ e^-=2(Z\d X-X^+X^- Z^2)=2e^Q(\d v\d X-e^Q Y \d^2 v ),
\eea
where $Y\equiv X^+ X^-$. 
So all solutions\footnote{in the linear dilaton vacua} for all generalized dilaton gravity models obey a generalized Birkhoff theorem, in the sense that all solutions exhibit a Killing vector $\partial_v$. The solution space is parameterized by two constants of integration. The one coming from the integration of \eqref{conserve} is non-trivial, while the one coming from \eqref{q} is trivial and can be fixed by a choice of units.
\subsection{Back to Second order formalism}
First we separate out the torsion-free part of the spin-connection. To do that we notice
\bea 
\star T_a=\star \( \d e_a+\epsilon_a^{~b} \omega \wedge e_b\)=\star \d e_a+\epsilon_a^{~b} \omega^c\star (e_c\wedge e_b).
\eea 
Using 
\bea 
&& e_a\wedge e_b=e_{a\mu}e_{b \nu}\d x^\mu \wedge \d x^\nu ,\\
&& \star e_a\wedge e_b=e_{a\mu}e_{b \nu}\star \d x^\mu \wedge \d x^\nu =e_{a\mu}e_{b \nu}\epsilon^{\mu\nu}=\epsilon_{ab}
\eea 
we get
\bea 
\star T_a=\star \d e_a+\epsilon_a^{~b}\omega^c\epsilon_{cb}=\star \d e_a-\omega_a.
\eea 
So we can rewrite the spin-connection as
\bea 
\omega=\omega^a e_a=\(\star\d e_a-\star T_a\)e^a=e^a\star \d e_a-e^a\star T_a.
\eea 
Then $\tilde{\omega}=e^a\star \d e_a$ is the torsion-free part which in terms of components is given by
\bea 
&&\star \d e_a=\partial_\mu(e_\nu)_a\epsilon^{\mu\nu},\quad \tilde{\omega}=e^a \partial_\mu(e_\nu)_a\epsilon^{\mu\nu}.
\eea 
Recall that the action in the first formalism is 
\bea 
I_{\text{gen}}\sim \int X \d \omega+\epsilon \mathcal{V}+X^a T_a.
\eea 
The first term can be manipulated as
\bea 
X\d\omega=-\d X\wedge \omega=-\d X\wedge (\tilde{\omega}-e^a\star T_a)=X \d \tilde{\omega}+\d X\wedge e^a \star T_a.
\eea 
Note that 
\bea 
\d \tilde{\omega}=\partial_{\mu}\omega_\nu \d x^\mu \wedge \d x^\nu \rightarrow -\frac{R}{2}\sqrt{-g}\d^2x,
\eea 
which is exactly the first term in the action \eqref{2dg}. It is obvious that 
\bea 
\epsilon \mathcal{V}(X,X^aX_a) \rightarrow \sqrt{-g}\mathcal{V}(X,X^aX_a) \d^2x.
\eea 
So the last thing to do is to remove $X^a$ with the help of equation of motion \eqref{e4}:
\bea 
&&T_a=-\frac{1}{2}\epsilon^{bc}e_b\wedge e_c \frac{\partial \mathcal{V}}{\partial X^a} \rightarrow \\
&&\star T_a=-\frac{1}{2}\epsilon^{bc}\epsilon_{bc}\frac{\partial \mathcal{V}}{\partial X^a} =\frac{\partial \mathcal{V}}{\partial X^a}
\eea 
and \eqref{e1}:
\bea 
&& \partial_\mu X+X^a \epsilon_a^{~b}e_{b\mu}=0 \rightarrow \\
&& \partial_{\mu}X+X^a e_a^\nu \epsilon_{\nu\mu}=0 \rightarrow \\
&& X^a=-e^a_\nu \epsilon^{\mu\nu}\partial_\mu X.
\eea 
So terms involved with $T^a$ are cancelled to each other:
\bea 
&& X^a T_a=\frac{1}{2}e^a_\nu \epsilon^{\mu\nu}\partial_{\mu} X\epsilon^{cb}e_c\wedge e_b \frac{\partial \mathcal{V}}{\partial X^a} \rightarrow e^a_\nu \epsilon^{\mu\nu}\partial_{\mu} X\frac{\partial \mathcal{V}}{\partial X^a} \sqrt{-g}\d^2 x,\\
&&\d X\wedge e^a\star T_a=\partial_\mu X e^a_\nu \frac{\partial \mathcal{V}}{\partial X^a} \d x^\mu \wedge \d x^\nu\rightarrow -e^a_\nu \epsilon^{\mu\nu}\partial_{\mu} X\frac{\partial \mathcal{V}}{\partial X^a} \sqrt{-g}\d^2 x.
\eea 
Then we arrive at the action in second order formalism
\bea 
-\frac{1}{2}\int \sqrt{-g}\(X R-2\mathcal{V}(X,-(\partial X)^2)\),
\eea 
where we have used
\bea 
X^aX_a=-(\partial  X)^2.
\eea 
Therefore we find that the $Q$ function is given by
\bea \label{Qfunc}
Q=\int^X U(y)\d y,
\eea 
and the conserved quantity \eqref{conserve} is given by
\bea 
&&C_0=e^Q Y+w,\quad Y=X^+ X^-,\quad w=\int^X e^Q V(y) \d y \label{wfun}\\
&&dC_0=e^Q \d Y+Y e^Q U(X)\d X+e^Q V \d X=0.
\eea 
Using this we can rewrite the metric as
\bea
\d s^2 &=& 2 e^Q \d v(\d X+(w(X)-C_0)\d v)\\
&=& 2\d v \d \tilde{X}+\xi(\tilde{X})\d v^2,
\eea 
where we have introduced
\bea 
\d \tilde{X}=\d X e^Q,\quad \xi(\tilde{X})=2e^Q(w-C_0). \label{xie}
\eea


\begin{thebibliography}{}
\bibitem{Li:2021lfo}
R.~Li, X.~Wang and J.~Wang,
``Island may not save the information paradox of Liouville black holes,''
Phys. Rev. D \textbf{104}, no.10, 106015 (2021)
doi:10.1103/PhysRevD.104.106015
[arXiv:2105.03271 [hep-th]].
\bibitem{Bousso:2022ntt}
R.~Bousso, X.~Dong, N.~Engelhardt, T.~Faulkner, T.~Hartman, S.~H.~Shenker and D.~Stanford,
``Snowmass White Paper: Quantum Aspects of Black Holes and the Emergence of Spacetime,''
[arXiv:2201.03096 [hep-th]].
\bibitem{Page:1993wv}
D.~N.~Page,
``Information in black hole radiation,''
Phys. Rev. Lett. \textbf{71}, 3743-3746 (1993)
doi:10.1103/PhysRevLett.71.3743
[arXiv:hep-th/9306083 [hep-th]].
\bibitem{Hawking:1975vcx}
S.~W.~Hawking,
``Particle Creation by Black Holes,''
Commun. Math. Phys. \textbf{43}, 199-220 (1975)
[erratum: Commun. Math. Phys. \textbf{46}, 206 (1976)]
doi:10.1007/BF02345020
\bibitem{Penington:2019npb}
G.~Penington,
``Entanglement Wedge Reconstruction and the Information Paradox,''
JHEP \textbf{09}, 002 (2020)
doi:10.1007/JHEP09(2020)002
[arXiv:1905.08255 [hep-th]].
\bibitem{Almheiri:2019psf}
A.~Almheiri, N.~Engelhardt, D.~Marolf and H.~Maxfield,
``The entropy of bulk quantum fields and the entanglement wedge of an evaporating black hole,''
JHEP \textbf{12}, 063 (2019)
doi:10.1007/JHEP12(2019)063
[arXiv:1905.08762 [hep-th]].
\bibitem{Almheiri:2019hni}
A.~Almheiri, R.~Mahajan, J.~Maldacena and Y.~Zhao,
``The Page curve of Hawking radiation from semiclassical geometry,''
JHEP \textbf{03}, 149 (2020)
doi:10.1007/JHEP03(2020)149
[arXiv:1908.10996 [hep-th]].
\bibitem{Engelhardt:2014gca}
N.~Engelhardt and A.~C.~Wall,
``Quantum Extremal Surfaces: Holographic Entanglement Entropy beyond the Classical Regime,''
JHEP \textbf{01}, 073 (2015)
doi:10.1007/JHEP01(2015)073
[arXiv:1408.3203 [hep-th]].
\bibitem{Penington:2019kki}
G.~Penington, S.~H.~Shenker, D.~Stanford and Z.~Yang,
``Replica wormholes and the black hole interior,''
JHEP \textbf{03}, 205 (2022)
doi:10.1007/JHEP03(2022)205
[arXiv:1911.11977 [hep-th]].
\bibitem{Almheiri:2019qdq}
A.~Almheiri, T.~Hartman, J.~Maldacena, E.~Shaghoulian and A.~Tajdini,
``Replica Wormholes and the Entropy of Hawking Radiation,''
JHEP \textbf{05}, 013 (2020)
doi:10.1007/JHEP05(2020)013
[arXiv:1911.12333 [hep-th]].
\bibitem{Sully:2020pza}
J.~Sully, M.~V.~Raamsdonk and D.~Wakeham,
``BCFT entanglement entropy at large central charge and the black hole interior,''
JHEP \textbf{03}, 167 (2021)
doi:10.1007/JHEP03(2021)167
[arXiv:2004.13088 [hep-th]].
\bibitem{Chen:2020uac}
H.~Z.~Chen, R.~C.~Myers, D.~Neuenfeld, I.~A.~Reyes and J.~Sandor,
``Quantum Extremal Islands Made Easy, Part I: Entanglement on the Brane,''
JHEP \textbf{10}, 166 (2020)
doi:10.1007/JHEP10(2020)166
[arXiv:2006.04851 [hep-th]].
\bibitem{Chen:2020hmv}
H.~Z.~Chen, R.~C.~Myers, D.~Neuenfeld, I.~A.~Reyes and J.~Sandor,
``Quantum Extremal Islands Made Easy, Part II: Black Holes on the Brane,''
JHEP \textbf{12}, 025 (2020)
doi:10.1007/JHEP12(2020)025
[arXiv:2010.00018 [hep-th]].
\bibitem{Suzuki:2022xwv}
K.~Suzuki and T.~Takayanagi,
``BCFT and Islands in Two Dimensions,''
[arXiv:2202.08462 [hep-th]].
\bibitem{Grumiller:2002nm}
D.~Grumiller, W.~Kummer and D.~V.~Vassilevich,
``Dilaton gravity in two-dimensions,''
Phys. Rept. \textbf{369}, 327-430 (2002)
doi:10.1016/S0370-1573(02)00267-3
[arXiv:hep-th/0204253 [hep-th]].
\bibitem{Grumiller:2006rc}
D.~Grumiller and R.~Meyer,
``Ramifications of lineland,''
Turk. J. Phys. \textbf{30}, 349-378 (2006)
[arXiv:hep-th/0604049 [hep-th]].
\bibitem{Iyer:1994ys}
V.~Iyer and R.~M.~Wald,
``Some properties of Noether charge and a proposal for dynamical black hole entropy,''
Phys. Rev. D \textbf{50}, 846-864 (1994)
doi:10.1103/PhysRevD.50.846
[arXiv:gr-qc/9403028 [gr-qc]].

\bibitem{Almheiri:2020cfm}
A.~Almheiri, T.~Hartman, J.~Maldacena, E.~Shaghoulian and A.~Tajdini,
``The entropy of Hawking radiation,''
Rev. Mod. Phys. \textbf{93}, no.3, 035002 (2021)
doi:10.1103/RevModPhys.93.035002
[arXiv:2006.06872 [hep-th]].

\bibitem{Krishnan:2020fer}
C.~Krishnan,
``Critical Islands,''
JHEP \textbf{01}, 179 (2021)
doi:10.1007/JHEP01(2021)179
[arXiv:2007.06551 [hep-th]].
\bibitem{Caceres:2020jcn}
E.~Caceres, A.~Kundu, A.~K.~Patra and S.~Shashi,
``Warped information and entanglement islands in AdS/WCFT,''
JHEP \textbf{07}, 004 (2021)
doi:10.1007/JHEP07(2021)004
[arXiv:2012.05425 [hep-th]].
\bibitem{Geng:2021wcq}
H.~Geng, Y.~Nomura and H.~Y.~Sun,
``Information paradox and its resolution in de Sitter holography,''
Phys. Rev. D \textbf{103}, no.12, 126004 (2021)
doi:10.1103/PhysRevD.103.126004
[arXiv:2103.07477 [hep-th]].
\bibitem{Geng:2021iyq}
H.~Geng, S.~L\"ust, R.~K.~Mishra and D.~Wakeham,
``Holographic BCFTs and Communicating Black Holes,''
jhep \textbf{08}, 003 (2021)
doi:10.1007/JHEP08(2021)003
[arXiv:2104.07039 [hep-th]].
\bibitem{Gautason:2020tmk}
F.~F.~Gautason, L.~Schneiderbauer, W.~Sybesma and L.~Thorlacius,
``Page Curve for an Evaporating Black Hole,''
JHEP \textbf{05}, 091 (2020)
doi:10.1007/JHEP05(2020)091
[arXiv:2004.00598 [hep-th]].
\bibitem{Hartman:2020swn}
T.~Hartman, E.~Shaghoulian and A.~Strominger,
``Islands in Asymptotically Flat 2D Gravity,''
JHEP \textbf{07}, 022 (2020)
doi:10.1007/JHEP07(2020)022
[arXiv:2004.13857 [hep-th]].
\bibitem{Hollowood:2020cou}
T.~J.~Hollowood and S.~P.~Kumar,
``Islands and Page Curves for Evaporating Black Holes in JT Gravity,''
JHEP \textbf{08}, 094 (2020)
doi:10.1007/JHEP08(2020)094
[arXiv:2004.14944 [hep-th]].
\bibitem{Goto:2020wnk}
K.~Goto, T.~Hartman and A.~Tajdini,
``Replica wormholes for an evaporating 2D black hole,''
JHEP \textbf{04}, 289 (2021)
doi:10.1007/JHEP04(2021)289
[arXiv:2011.09043 [hep-th]].
\bibitem{Chen:2020jvn}
H.~Z.~Chen, Z.~Fisher, J.~Hernandez, R.~C.~Myers and S.~M.~Ruan,
``Evaporating Black Holes Coupled to a Thermal Bath,''
JHEP \textbf{01}, 065 (2021)
doi:10.1007/JHEP01(2021)065
[arXiv:2007.11658 [hep-th]].
\bibitem{Wang:2021mqq}
X.~Wang, R.~Li and J.~Wang,
``Page curves for a family of exactly solvable evaporating black holes,''
Phys. Rev. D \textbf{103}, no.12, 126026 (2021)
doi:10.1103/PhysRevD.103.126026
[arXiv:2104.00224 [hep-th]].
\bibitem{Almheiri:2019psy}
A.~Almheiri, R.~Mahajan and J.~E.~Santos,
``Entanglement islands in higher dimensions,''
SciPost Phys. \textbf{9}, no.1, 001 (2020)
doi:10.21468/SciPostPhys.9.1.001
[arXiv:1911.09666 [hep-th]].
\bibitem{Hashimoto:2020cas}
K.~Hashimoto, N.~Iizuka and Y.~Matsuo,
``Islands in Schwarzschild black holes,''
JHEP \textbf{06}, 085 (2020)
doi:10.1007/JHEP06(2020)085
[arXiv:2004.05863 [hep-th]].
\bibitem{Wang:2021woy}
X.~Wang, R.~Li and J.~Wang,
``Islands and Page curves of Reissner-Nordstr\"om black holes,''
JHEP \textbf{04}, 103 (2021)
doi:10.1007/JHEP04(2021)103
[arXiv:2101.06867 [hep-th]].
\bibitem{Yu:2021cgi}
M.~H.~Yu and X.~H.~Ge,
``Islands and Page curves in charged dilaton black holes,''
Eur. Phys. J. C \textbf{82}, no.1, 14 (2022)
doi:10.1140/epjc/s10052-021-09932-w
[arXiv:2107.03031 [hep-th]].
\bibitem{Ahn:2021chg}
B.~Ahn, S.~E.~Bak, H.~S.~Jeong, K.~Y.~Kim and Y.~W.~Sun,
``Islands in charged linear dilaton black holes,''
Phys. Rev. D \textbf{105}, no.4, 046012 (2022)
doi:10.1103/PhysRevD.105.046012
[arXiv:2107.07444 [hep-th]].
\bibitem{Karananas:2020fwx}
G.~K.~Karananas, A.~Kehagias and J.~Taskas,
``Islands in linear dilaton black holes,''
JHEP \textbf{03}, 253 (2021)
doi:10.1007/JHEP03(2021)253
[arXiv:2101.00024 [hep-th]].
\bibitem{Lu:2021gmv}
Y.~Lu and J.~Lin,
``Islands in Kaluza\textendash{}Klein black holes,''
Eur. Phys. J. C \textbf{82}, no.2, 132 (2022)
doi:10.1140/epjc/s10052-022-10074-w
[arXiv:2106.07845 [hep-th]].
\bibitem{Krishnan:2020oun}
C.~Krishnan, V.~Patil and J.~Pereira,
``Page Curve and the Information Paradox in Flat Space,''
[arXiv:2005.02993 [hep-th]].
\bibitem{Alishahiha:2020qza}
M.~Alishahiha, A.~Faraji Astaneh and A.~Naseh,
``Island in the presence of higher derivative terms,''
JHEP \textbf{02}, 035 (2021)
doi:10.1007/JHEP02(2021)035
[arXiv:2005.08715 [hep-th]].
\bibitem{Anegawa:2020ezn}
T.~Anegawa and N.~Iizuka,
``Notes on islands in asymptotically flat 2d dilaton black holes,''
JHEP \textbf{07}, 036 (2020)
doi:10.1007/JHEP07(2020)036
[arXiv:2004.01601 [hep-th]].
\bibitem{He:2021mst}
S.~He, Y.~Sun, L.~Zhao and Y.~X.~Zhang,
``The universality of islands outside the horizon,''
[arXiv:2110.07598 [hep-th]].
\bibitem{Callan:1992rs}
C.~G.~Callan, Jr., S.~B.~Giddings, J.~A.~Harvey and A.~Strominger,
``Evanescent black holes,''
Phys. Rev. D \textbf{45}, no.4, R1005 (1992)
doi:10.1103/PhysRevD.45.R1005
[arXiv:hep-th/9111056 [hep-th]].

\bibitem{Witten:2020wvy}
E.~Witten,
``Matrix Models and Deformations of JT Gravity,''
Proc. Roy. Soc. Lond. A \textbf{476}, no.2244, 20200582 (2020)
doi:10.1098/rspa.2020.0582
[arXiv:2006.13414 [hep-th]].

\bibitem{Cruz:1997nj}
J.~Cruz, J.~Navarro-Salas, C.~F.~Talavera and M.~Navarro,
``Conformal and non-conformal symmetries in 2-D dilaton gravity,''
Phys. Lett. B \textbf{402}, 270-275 (1997)
doi:10.1016/S0370-2693(97)00458-9
[arXiv:hep-th/9606097 [hep-th]].
\bibitem{Mann:1993rf}
R.~B.~Mann,
``Liouville black holes,''
Nucl. Phys. B \textbf{418}, 231-256 (1994)
doi:10.1016/0550-3213(94)90246-1
[arXiv:hep-th/9308034 [hep-th]].
\bibitem{Katanaev:1996ni}
M.~O.~Katanaev, W.~Kummer and H.~Liebl,
``On the completeness of the black hole singularity in 2-d dilaton theories,''
Nucl. Phys. B \textbf{486}, 353-370 (1997)
doi:10.1016/S0550-3213(96)00624-4
[arXiv:gr-qc/9602040 [gr-qc]].
\end{thebibliography}
\end{document}